\documentclass[preprint,10pt]{aastex}
\usepackage{lscape}
\usepackage{epstopdf}
\usepackage{epsfig,natbib}
\usepackage{graphicx}

\usepackage{verbatim}
\shortauthors{Tellis \& Marcy}
\shorttitle{Optical Laser SETI}

\begin{document}
\pagenumbering{arabic}

\title{A Search for Optical Laser Emission Using Keck HIRES{$\dagger$}}

\author{ Nathaniel~K.~Tellis\altaffilmark{1,2}, Geoffrey~W.~Marcy\altaffilmark{1,3}}                
\email{nate.tellis@gmail.com}                                                                                               
\altaffiltext{1}{Astronomy Department, University of California, Berkeley, Berkeley, CA 94720}  
\altaffiltext{2}{Department of Physics, McGill University, Montreal, QC H3A 0G4, Canada}  
\altaffiltext{3}{Alberts SETI Chair, Berkeley SETI Research Center}  
                                                                                                
\altaffiltext{$\dagger$}{We dedicate this work to the memory of Charles H. Townes, inventor of the laser
  and pioneer of optical SETI.} 

\begin{abstract}
We present a search for laser emission coming from point sources in
the vicinity of 2796 stars, including 1368 {\em Kepler} Objects of
Interest (KOIs) that host one or more exoplanets.  We search for
extremely narrow emission lines in the wavelength region between 3640 and 7890
\AA \ using the Keck 10-meter telescope and spectroscopy with high
resolution ($\lambda/\Delta \lambda$ = 60,000). Laser emission lines
coming from non-natural sources are distinguished from natural
astrophysical sources by being monochromatic and coming from an
unresolved point in space.  We search for laser emission located 2-7
arcsec from the 2796 target stars.  The detectability of laser
emission is limited by Poisson statistics of the photons and scattered
light, yielding a detection threshold flux of $\sim10^{-2}$ photons
m$^{-2}$ s$^{-1}$ for typical {\em Kepler} stars and 1 photon
m$^{-2}$s$^{-1}$ for solar-type stars within 100
light-years. Diffraction-limited lasers having a 10-meter aperture can
be detected from 100 light-years away if their power exceeds 90 W, and
from 1000 light-years away ({\it Kepler} planets), if their power
exceeds 1 kW (from lasers located 60-200 AU, and 2000-7000 AU from the
nearby and {\it Kepler} stars, respectively).  We did not find any
such laser emission coming from any of the 2796 target stars.  We
discuss the implications for the search for extraterrestrial
intelligence (SETI).
\end{abstract}


\clearpage

\section{Introduction}

During the past 65 years, the search for extraterrestrial intelligence
(SETI) has proceeded with innovative attempts to detect the generation
of light from civilizations residing elsewhere in the universe, e.g.,
\citep{Cocconi_Morrison1959, Drake1961, Tarter2001, Siemion2013}.
Searches have been conducted using a variety of wavelengths, most
prominently with radio telescopes sensitive to centimeter wavelengths with
Arecibo (L band), at 1/10 meter wavelengths with the
Greenbank radio telescope, and at centimeter wavelengths with the
Allen Telescope Array \citep{Drake1961, Werthimer2001, Korpela2011, Tarter2011}, some targeting arbitrary stars and galaxies and
recently aimed toward known exoplanets \citep{Siemion2013}. Searches at
optical wavelengths (``OSETI'') have also been conducted, both for
continuously transmitting lasers and for sub-$\mu$s-duration light
pulses \citep{Reines2002, Wright2001, Howard2004, Stone2005, Howard2007}.  To date, no
convincing evidence of other technological civilizations has been
found.

New SETI efforts have begun that employ new wavelengths and achieve
greater sensitivity. One search was conducted at mid-infrared
wavelengths to detect the waste thermal emission from the vast
machinery of advanced civilizations, often called ``Dyson spheres''
\citep{Wright2014_J}.  Some SETI efforts involve detecting the dimming
of starlight as planet-sized technological constructs pass in front of
stars \citep{Walkowicz2014, Forgan2013}.  Future SETI efforts will
take advantage of the next generation of large radio telescopes,
including the Square Kilometre Array \citep{Siemion2014}.  There are
also proposals to broadcast (rather than just receive) bright beacons
as Galactic marquees to advertise our human presence. But
international agreements are needed before we compose and transmit
brilliant messages moving irreversibly at the speed of light, with
unknown consequences \citep{Shostak2013, Brin2014}.  Recently, there
has been increased interest in searches for extraterrestrial optical
and near-infrared lasers, including those emitting short-duration
pulses or periodicities \citep{Howard2004, Howard2007, Korpela2011,
  Drake2010, Mead2013, Covault2013, Leeb2013, Gillon2014,
  Wright2014_S}.

The advantages of interstellar communication by optical and IR lasers
include the ease in producing high intensity, diffraction-limited
beams to transmit over Galactic distances, and achieving relative privacy
with high data rates exceeding 10$^{\rm12}$ bits per second
\citep{Townes_Schwartz1961, Wright2001}. Military and commercial lasers have
demonstrated continuous-wave power near 30 kW, and 100 kW power is
planned \citep{lockheed2014}. Astronomers inadvertently shoot laser
beams toward interesting astronomical objects, including exoplanets
and the Galactic center, by employing laser guide star adaptive optics
on large telescopes with typical power of $\sim$7 W. NASA has
demonstrated the use of pulsed infrared lasers for Earth-to-Moon
communication at a data rate of 622 Megabits per second
\citep{Buck2013}. Optical and IR lasers would also serve for
satellite-satellite data transfer. For communication between more
distant celestial bodies, optical and IR lasers may be especially
useful due to the tight beam, offering enhanced energy efficiency and
reduced eavesdropping. In this current paper, we search for laser emission coming from 
spatial regions separated from stars by tens to hundreds of AUs 
in projection, where the laser light does not have to compete with                                      
the starlight. Section \ref{sec:sens} contains a further 
discussion of the relative strengths of lasers needed for detection by our method. 

An advanced civilization might emit many lasers of varying beam sizes,
constrained by available optical technology, resulting in unavoidable
spill-over at the receiver. The laser light would continue propagating
in the original direction, past the intended receiver. The filling
factor of such beams in the Milky Way Galaxy is completely unknown.
More purposefully, an advanced civilization seeking to indicate its
presence to nearby habitable planets might use lasers to offer a
bright and unambiguous beacon of its presence.  Natural astrophysical
conditions exist that allow gas to lase in the near-infrared.  But the
only observed astrophysically pumped lasers at optical wavelengths occur in
non-LTE conditions that produce population inversions in neutral
oxygen, causing extraordinary emission at [O I] 8446 \AA,
\citep{Johansson2004, Johansson2005}. The presence of a bright,
unresolved emission line at any other optical wavelength would be
worthy of further attention, including the possibility of a
non-astrophysical source.

The past 20 years of discoveries of exoplanets motivate a new
perspective about SETI. To date, over 1500 exoplanets have been
confirmed with accurate orbits, and another 3300 planet candidates
have been identified by {\it Kepler} as likely real but still need
confirmation at the 99\% confidence level \citep{Rowe2014a, Rowe2014b,
  Mullally2014, Wright2011_J}.  Stars with known exoplanets make
excellent targets for SETI searches, and such planetary systems have
already been surveyed for technological transmissions at radio
wavelengths \citep{Siemion2013, Gautam2014}.  Planetary systems found
by {\it Kepler} are particularly valuable for SETI searches because
their edge-on orbital planes enhance the probability of our detecting
spillover transmissions  between planets and satellites within that planetary
system\footnote{The powers needed to transmit within a
planetary system are clearly dwarfed by those needed over interstellar 
distances.}.

However, {\it Kepler} showed that over 50\% of both solar-type stars
and M-dwarf stars have planets within 1 AU of the host star
\citep{Petigura2013, Dressing2013}. Planets smaller than 1.5
Earth-radii are common and often harbor a largely rocky interior
\citep{Weiss2014, Rogers2014, Wolfgang2014, Marcy2014}.  Thus, all
FGKM-type stars have comparable value as targets for SETI searches,
whether or not they have known exoplanets, as the majority of stars
harbor planets of 1-2 Earth-radii within a few AU \citep{Howard2012,
  Petigura2013}. Only binary stars separated by less than a few AU are
probably poor sites for SETI searches, due to a lack of stable
planetary orbits. Otherwise, planets with liquid water may permit
progressively complex organic chemistry toward the nucleotides and
duplication of RNA within fatty acid proto-membranes
\citep{Adamala2013, Szostak2012}.

In Section~\ref{sec:hires} we describe the selection of target stars and
the spectroscopic instrumentation to detect optical laser emission lines. 
In Section~\ref{sec:algorithm} we describe our
laser line detection algorithm.  In Section~\ref{sec:results} we
present the results of our search for laser lines and in section
\ref{sec:sens} we specify the detection thresholds of laser emission
and the strength of transmitters to which we are sensitive. Finally,
in Section~\ref{sec:Disc} we offer a discussion of the results in the
context of the extraterrestrial transmission of laser beacons.

\section{Target Stars and Spectroscopic Search for Laser Emission}
\label{sec:hires}

\subsection{Target Stars}
\label{sec:targets}

All 2796 target stars in this search for laser emission were stars for
which high resolution spectra had already been obtained at the Keck
Observatory as part of a study of their exoplanets.  There were two
populations of target stars drawn from our two broad exoplanet
programs. The first population of targets stems from the California
Planet Search (CPS) that continue to make repeated Doppler
measurements of over 3000 FGKM main sequence stars brighter than Vmag
= 8.5 and northward of declination -25 deg, ongoing for the past 10-20
years, using both the Lick and Keck Observatories \citep{Fischer2014,
  Howard2014, Marcy2008, Wright2011_J}. Numerous papers have been
written about the detections and properties of the exoplanets found
among these 3000 stars, e.g., \citep{Marcy2008, Johnson2011,
  Howard2014}. Roughly 10\% of these stars have known planets detected
by the RV method \citep{Cumming2008, Howard2010}.  All M dwarfs
brighter than Vmag=11 are also being followed typically with several
Doppler measurements per year.  The CPS exoplanet survey continues
primarily with the Keck 1 telescope now that the iodine cell at the
Lick Observatory 3-m Shane telescope has been retired.

For the second population of targets, we had taken spectra with the
Keck 1 HIRES spectrometer of all 1100 {\em Kepler} Objects of Interest
(KOI) that are brighter than Kepmag=14.2. 
We also have spectra of another 200 fainter KOIs that harbor 4, 5, or
6 transiting planets. These 1300 KOIs are predominantly FGKM main
sequence stars identified by {\em Kepler} as likely harboring one or
more planets, and over 90\% of them have real planets, as opposed to
false positives \citep{Morton2011}. Nearly all of the multi-transiting
planet systems are already confirmed as real planets
\citep{Lissauer2014, Rowe2014a}.  A list of these KOIs and measurements
of each star's effective temperature (Teff), surface gravity (log g),
iron abundance ([Fe/H], and projected rotational velocity (Vsini) are
provided by \citep{Petigura2015, Howard2015}.

Most spectra were taken with an exposure meter that stopped the
exposure when a pre-set number of photons was received per pixel, with
the typical goal of achieving a signal-to-noise ratio in the reduced
spectrum of 100-200.  The resulting exposure times varied from 1 to 45
minutes to accommodate the factor of $\sim$100 range in brightnesses
of the target stars, which in turn depend on their intrinsic
luminosity in visible light and their distance.  The entrance slit of
the HIRES spectrometer was oriented perpendicular to the horizon using
an optical image rotator.  Thus, the orientation of the slit was
neither N-S nor E-W in equatorial coordinates, but depended on the
Hour Angle at the time of observation and the Declination.

These spectra of 3000 main sequence and subgiant stars offer a fresh
opportunity to search for pulsed and continuous optical laser
emission.  We carried out a similar search for unresolved laser lines
in HIRES spectra on a small subset of the stars surveyed here
\citep{Reines2002}. Our present sample of stars is much larger,
including KOIs, and the laser-line search algorithm is greatly
improved, operating on raw CCD images rather than reduced spectra.

The first target population of nearby FGKM stars are brighter than
Vmag=11, with most brighter than Vmag=8.5.  The typical exposure times
to obtain the spectra were 1 minute at Vmag=7 and 8 minutes at
Vmag=10, varying by factors up to three depending on clouds and
seeing.  The {\em Kepler} target stars typically have ($Vmag = 11-13$)
forcing exposure times of 20 - 45 minutes, depending on brightness,
clouds, and seeing. The longer exposure times of the fainter
population of stars permits detection of laser emission arriving with
lower flux at the Keck telescope to achieve requisite integrated
photons for detection (see Section~\ref{sec:sens}).  We also obtained
spectra of a small number of nearby galaxies, supernovae, and
planetary nebulae motivated by isolated projects of special interest.

In this work, we only examined those spectra (of the two populations
of stars) obtained with the 14x0.87 arcsec, ``C2'', entrance slit at
the HIRES spectrometer. The long 14 arcsec slit enables examination of
the region angularly near the star (``sky'') to detect laser emission
with no competition from the star's light. No filters were used during
the exposures, notably the KV370 filter that removes the light
shortward of 370 nm was not used.  A small amount of UV light from
second-order diffraction off the cross disperser leaks onto the CCD
longward of nominal 600 nm (300 nm, second order).  Many observations
used here had the iodine cell in place, but this does not affect the
search for laser lines displaced spatially from the star.

\begin{figure}
  \begin{center}
    \leavevmode
      \includegraphics[width=\textwidth]{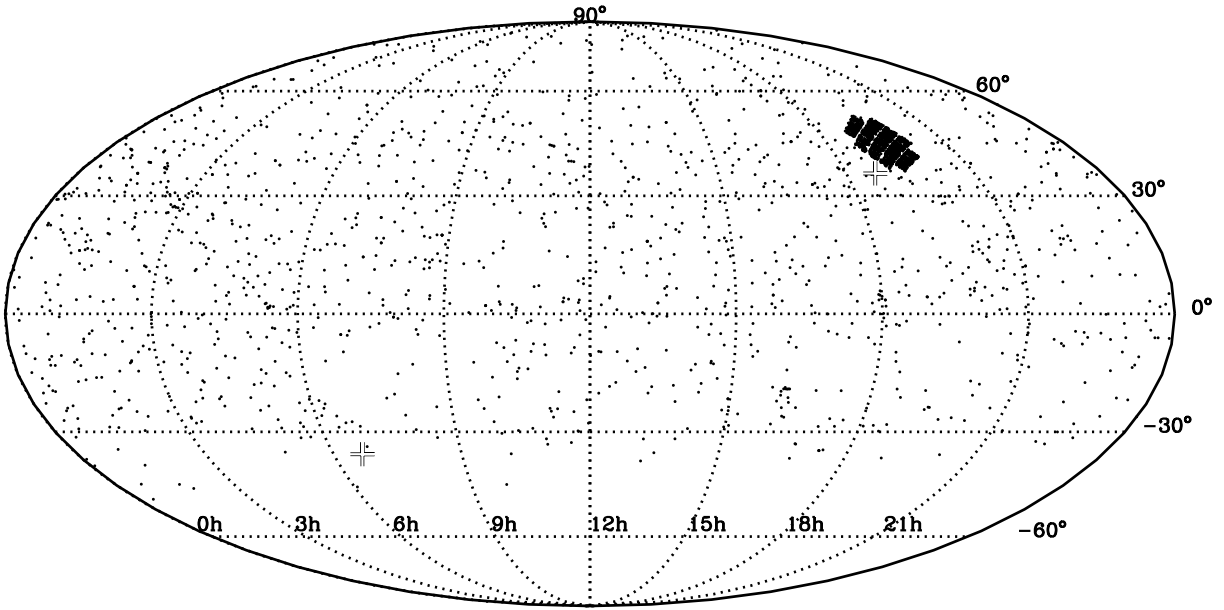}
       \caption[Coordinates]{The celestial coordinates, RA and DEC, of all 2796 target stars surveyed spectroscopically at
         the Keck Observatory for laser emission in the wavelength
         range, 3640-7890 \AA.  The faint crosses show the location of
       the apex, and anti-apex, of solar motion}
     \label{fig:targetRAandDEC}
  \end{center}
\end{figure}

We have a total of 14,380 spectra of 2796 stars obtained with that C2
decker, with many stars having multiple spectra of them taken between
2004 and 2014. The 2796 target stars are composed of 1,368 KOIs with
the remaining targets being nearby FGKM stars on the CPS
exoplanet program, and some additional planet searching targets.  Figure \ref{fig:targetRAandDEC} shows the location
of all 2796 target stars for this SETI search, in the domain of their
coordinates, RA and DEC. The figure shows that target stars are
located at all RA and are mostly north of DEC $>$ -30 deg.  There is a
concentration of 1368 targets in the  {\em Kepler} field of view between RA =
19-20 and DEC = 35-50 deg.
\\
\subsection{HIRES Spectrometer setup}
\label{sec:HIRES_setup}

All 14,380 spectra were obtained with the HIRES spectrometer on Keck 1
\citep{Vogt1994}.  The spectra span wavelengths 3640 and 7890 \AA \,
split among 49 spectral orders that fall on three 2048 by 4096 pixel
CCDs.  The separation of the spectral orders varies from 6 arcsec in
the near UV to 43 arcseconds in the far red \citep{Griest2010,
  Vogt1994}. In order to diminish the CCD readout time, three CCD
pixels are binned in the spatial dimension on chip.  The spectral
resolution (FWHM, $\Delta \lambda$, of the instrumental profile) is $\lambda/\Delta \lambda$=60,000, i.e., 5 km s${-1}$). Each pixel spans 1.3 km
s${-1}$ in the wavelength direction and 0.38 arcsec (after binning) in
the spatial direction.

\begin{figure}
  \begin{center}
    \leavevmode
      \includegraphics[width=\textwidth]{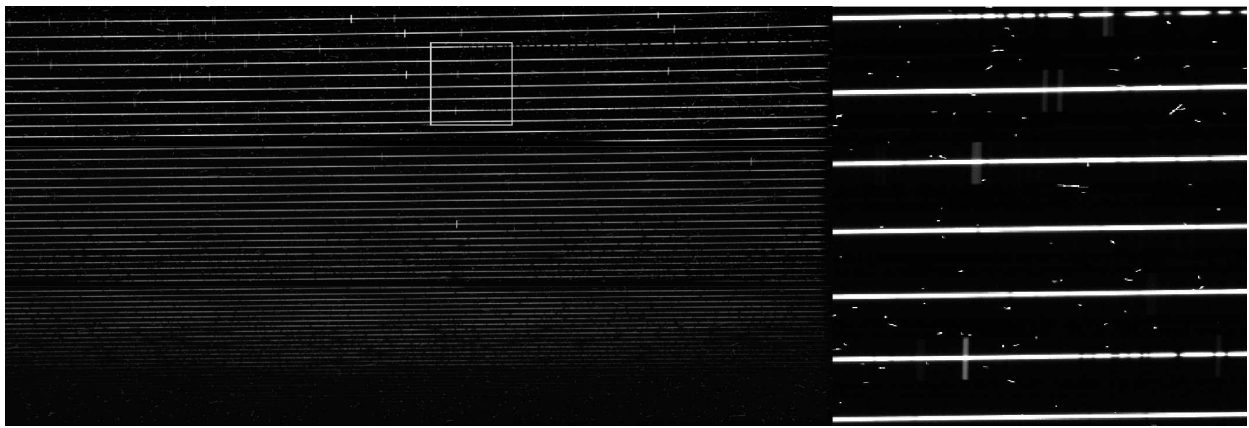}
       \caption[HIRES Spectrum]{A representative raw echelle spectrum analyzed here for laser lines.  The
         horizontal stripes are the spectral orders containing roughly
         100 \AA \ segments of the optical spectrum, each displaced
         vertically from its neighbors by typically 40 pixels,
         corresponding to $\sim$ 15 arcsec projected on the
         sky. Wavelength increases from left to right along each
         order, and from bottom to top among orders. The boxed region
         is shown in detail to the right. Note the night sky emission
         lines that span 14 arcseconds in the spatial direction
         perpendicular to each spectral order. We survey for laser
         emission 2-7 arcsec above and below each spectral order at
         each wavelength from 3640-7980 \AA. }
     \label{fig:keckspec}
  \end{center}
\end{figure}

The search for laser lines was performed on the raw CCD images, not
the reduced spectra.  We specifically searched for laser emission in
the spatial region between 2-7 arcsec from the star in both directions
along with length of the slit.  All raw CCD images of the spectra use
here are available to the public at the Keck Observatory Archive.
We show one example of the raw CCD image in
Figure \ref{fig:keckspec}. The use of a long 14x0.87 arcsec
C2 entrance slit placed a spectrum of the ``sky'' on either side of
the star, making it possible to probe that sky for laser emission
coming from nearby region of the star with little contamination from
the starlight itself.  In this paper, "rows" denote
pixels along the direction of the dispersion of the spectrum and
``columns'' denotes pixels along the
spatial direction perpendicular to dispersion, as shown in Figure \ref{fig:keckspec}.

\subsection{Laser Emission Properties}
\label{sec:laser}

We define properties of ``laser emission'' for this search
as sources located so far away that
they would be spatially unresolved as viewed by the Keck telescope on
Mauna Kea, corresponding to an angular size smaller then $\sim$0.8
arcsec, which is set by the atmospheric seeing.  Any laser-emitting machine 
having a projected size of $L$ would be unresolved if located at a
distance greater than, $d > 2\times 10^5 L$.  Machines with sizes
measured in tens of meters would obviously be unresolved at the
distances of the nearest stars. We restrict our search to such
spatially unresolved sources.

We also define ``laser emission'' to be emission with a linewidth
negligible compared to the resolution of the spectra we obtained with
Keck HIRES, $R = \lambda/\Delta\lambda =
6\times10^4$.  Here, $\Delta\lambda$ is the FWHM of the typical
instrumental profile of HIRES as used here. For comparison, typical
modern continuous wave (CW) lasers have linewidths limited by the
coherence length, the Doppler effect of the atoms, and the mechanical
stability of the laser cavity, yielding linewidths smaller than 1 GHz,
i.e. monochromatic with $\lambda/\Delta\lambda > 1\times10^6$, rarely
reaching the fundamental quantum Schawlow-Townes laser linewidth of
under 1 kHz.  (Pulsed dye lasers have much broader linewidths.)  For
reference, HeNe lasers with wavelength 632.8 nm and a linewidth of
1MHz ($\Delta\lambda = 1\times10^{-6}$ nm) are commercially available
and unresolvable by the HIRES spectrometer. As another reference, a
typical CW laser guide star for adaptive optics systems has an output
power of 20 W and operates at the blueward sodium D-line at
a wavelength, $\lambda$=589.1 nm, and a linewidth of 5 MHz, many orders of
magnitude narrower than resolvable by HIRES. Thus, a necessary
condition to be deemed ``laser
emission'' here is a line width narrower than HIRES can resolve.  
Any actual extraterrestrial lasers with
linewidths greater than the resolution of HIRES will not be identified
in this search. Such broadband signals could arrive from
technological civilizations, but this current work will not detect it.   
Any such technological broadband emission is more difficult to distinguish
from the kinematically or thermally broadened lines from naturally
occurring astrophysical sources.

Any laser source spatially separated from the star by more than the
seeing disk radius in the spatial direction will appear in the telescope focal plane as a
separate, unresolved image that will not overlap with
star light. We thus focus our attention here on portions
of the spatial profile separated from the star by the seeing
profile. As the seeing is typically 0.8 arcsec at optical wavelengths
at Mauna Kea, we are examining here the 4.5 arcsec of slit real estate
commonly called ``sky'', located more than 2 arcsec away from, and on
either side of, the stellar spectrum. A potential laser would be
effectively a point source in space and wavelength.  We expect it to
appear on the raw CCD image as a ``dot'', with a two-dimensional point spread
function (2-D PSF) shape corresponding to the seeing disk in the spatial direction and to the
spectral instrumental profile of HIRES in the wavelength
direction. Such dots would appear in the ``sky'' region of the
slit image where typically only a few photons hit, allowing the laser
to stand out against that faint background.  A pulsed laser would also
be detectable as a dot, provided the pulse duty cycle and pulse
energies produced a time-averaged power sufficient to yield the
threshold photons during an exposure, as described in Section~\ref{sec:sens}.

\subsection{Specifying the 2-D PSF of Candidate Laser Lines}
\label{sec:method}

We search for laser emission by examining the three ``raw'' CCD images
written by HIRES after each exposure. We do not use reduced
spectra. This allows us to search for laser emission ``dots''
consistent with the shape of the PSF both spatially and spectrally, as
defined above in Section~\ref{sec:laser}.  This two-dimensional search is
one key difference between this work and the earlier effort
\citep{Reines2002}, which made use of reduced spectra and therefore did
not benefit from spatial PSF information.  In addition, by retaining
the spatial information, we more easily distinguish unresolved
laser sources from spatially resolved sources such as nebulae,
extended galaxies, and night sky emission lines.

In broad outline, our algorithm treats each observation separately by
measuring its PSF in both the spatial and spectral directions as
described in Section~\ref{sec:algorithm}.  The code then steps
pixel-by-pixel along the ``sky'' pixels located spatially adjacent to
stellar spectral order and performs diagnostics on subimages to
identify laser line candidates.


The point spread function (PSF) is modeled as a 2-dimensional Gaussian
in the spatial and wavelength directions, as shown in Figure \ref{fig:interior}.
 The spatial and wavelength
full-width-half-maxima (FWHM) are set by measuring the average FWHM of
the spectral orders containing the stellar spectrum and the average
FWHM of the telluric night sky emission lines at various wavelengths,
respectively.  The FWHM in both directions varies by 5-10\% over the echelle
spectrum format due to the performance of HIRES optics over the entire
echelle field of view.   This variation plays only a minor role
in our criteria for goodness-of-fit of any detected laser lines due to
our relaxed standards for accepting candidate laser lines, as described below.

The FWHM of the spatial profile varies by 6$\%$, RMS, over the echelle
format in a given exposure due mostly to camera optics. Of course the spatial profile changes
greatly from hour to hour and night to night as atmospheric seeing
conditions vary. We account for this variability by measuring the
width of the stellar spectrum over the echelle format in each
exposure. We measure the spatial FWHM in 10 evenly spaced locations
within each spectral order. At each location we bin 11
columns \footnote{Columns containing outlier count levels, often due
to cosmic rays, are identified and ignored in the binning}. We then
fit a Gaussian to the resulting binned spectrum, and obtain a value
for the FWHM at the location. The result is a set of 10 FWHMs for each
spectral order. To get the local FWHM within an order, we linearly
interpolate between the ten points. The error imparted by the linear
interpolation is less than a percent, and is accounted for later in
Section \ref{sec:testing}.

The FWHM of the instrumental profile in the wavelength direction is
measured from night sky emission lines that are intrinsically
unresolved at our resolution of $R=6\times10^4$. We use the
atmospheric [OI] lines at 6300\AA \ and 5577\AA, which are present in
all observations, are quite bright, and vary in FWHM by less than a
few percent from night to night due to our vigilance focusing the
HIRES spectrometer. These night sky lines provide an acceptable proxy
to a delta function in wavelength, as their line widths are well below
the wavelength resolution achieved by HIRES \citep{Osterbrock1996}. To
characterize the FWHM in wavelength, the rows containing the night sky
lines are binned together (less those with significant contributions
from either the stellar spectrum or cosmic ray hits). A Gaussian is
then fit to the resulting binned vector. The average of these two
values for the FWHM - at 6300\AA \ and 5577\AA \ - provide an
approximation to the FWHM of the instrumental profile on a given
night. Some other atmospheric lines on the redder end of the spectrum
were considered for use, but most of them have greater variability in
their brightness and they change with time of day and
season. Nonetheless, the FWHM of the instrumental profile varies by
$\sim$10\% over the echelle format due to the normal optical
performance over the full field of view, and we account for this
variation in Section \ref{sec:testing}.

Rather than construct a 2-D PSF anew for each wavelength region, we
generate a small set of PSFs that vary in the spatial direction, as
observed. As we scan the CCD for laser emission, we draw from a
library of PSFs rather than generating them anew at each location.
While there is some loss of precision by drawing PSFs this way, we
tested the loss of detectability of laser emission that we incur and
found that the loss in sensitivity was marginally significant due to
our use of simplified PSFs.

\subsection{Location of the Stellar Orders}

We aim to search for laser emission that is spatially separated from
the stellar image.  Each spectral order on the CCD has a width set by
the size of the stellar image, set by the atmospheric seeing during
the exposure.  Thus, for each image we had to determine, to within one pixel, the
locations of the ridge and the width of each spectral order containing
the starlight. We aim to analyze only the ``sky-illuminated'' portions
of the CCD that have no stellar light but may contain laser light.  
We ignore, of course, those CCD pixels between the orders on which no
``sky'' light falls, as no laser light could hit them either.  That
is, the slit image does not extend over the entire CCD region between
each order for all wavelengths longward of 5500 \AA.

To locate the ridge and width of the orders containing stellar light
we first perform a 3x3 median filter on all three CCD images.  This
removes cosmic rays \footnote{Cosmic rays usually hit in one or a few
pixels, and only sometimes do they raise the counts in more than 4
pixels within a 3x3 region. Note that the 3x3 median filter was used 
exclusively for the location of the stellar ridge, and was not retained 
for the subsequent pixel-by-pixel laser search.}, ensuring that the ridge of the stellar
spectrum has the most photon counts along the order. To find the
ridge of each order we bin each set of 10 columns and identify the row
having the maximum value.

The locations of the orders vary by less than a pixel from observation
to observation due to our consistent optical set-up of the gratings in
the HIRES spectrometer and our accurate guiding of the star image on
the entrance slit.  We perform a constrained linear interpolation
between these ridge locations to establish the approximate location of
each spectral order.  We avoid searching for laser lines within $\pm
1.5 \times {\rm FWHM}_{spatial}$ from the ridge of the stellar order, to
avoid contamination from stellar continuum light.   Figure
\ref{fig:overlap} shows the geometry of the orders and the region
between them.

Shortward of approximately 5500 \AA, the slit images of the starlight
 overlap between the orders.  We use our determination of the spatial profile to
 anticipate such regions that must be avoided in the search for laser
 lines.  There simply are no ``sky'' pixels in between non-overlapping orders , so these regions of
 the CCD were ignored in the search for laser lines.  For spectral
 orders close enough spatially to each other that they have
 overlapping ``sky'', we do search for laser lines in that sky
 region, as shown in Figure \ref{fig:overlap}. 
In such cases, we cannot determine which order and which of the two corresponding
 wavelengths is associated with any laser line we may detect.

\subsection{Telluric Lines and Artifacts}

Terrestrial nightglow emission lines from the night sky (such as [OI]
5577.3 and 6300.3 \AA) are too bright to permit detection of laser
lines and they can fool our algorithms designed to search for laser
lines, especially on their edges. We also ignore a large artifact
in the raw HIRES images, a diagonal stripe that runs through the
center of the middle CCD known colloquially as the
``Meteor'' \footnote{the Meteor is actually scattered light dispersed
  twice by the spectrograph \citep{HIRESDATA}}.

On the other hand, faint night sky emission lines often are
sufficiently bright to fool our laser-line searching algorithms,
producing an apparent signal-to-noise ratio and partial shapes at
their ends ($\pm 7$\rq\rq{} from the center of the stellar spectrum).
Such faint night sky lines often meet the PSF criterion in the
wavelength dimension, and can marginally match the criterion in
one spatial direction, at the spatial edge of night sky line. This,
coupled with Poisson fluctuations in the brightness and in the
background count levels, can cause the ends of night sky lines to be
identified erroneously as candidate laser lines.

We ameliorate this problem in two ways. First, we compare pixel
locations to a catalog of OH and O$_2$ night sky lines at Mauna Kea
\citep{Osterbrock1996}. In order to not miss any, we convolve the CCD
images with a 2-D Gabor filter oriented in the direction of the night
sky lines, and omit results from wavelength values with large
spatially oriented signal. The latter technique is necessary, as the
laser search pipeline is sensitive enough to pick out faint night sky
lines, and not all lines are present in all observations.

\section{Identifying Laser Lines by S/N and $\chi^2$}
\label{sec:algorithm}

\begin{figure}
  \begin{center}
      \includegraphics[width=0.5\textwidth]{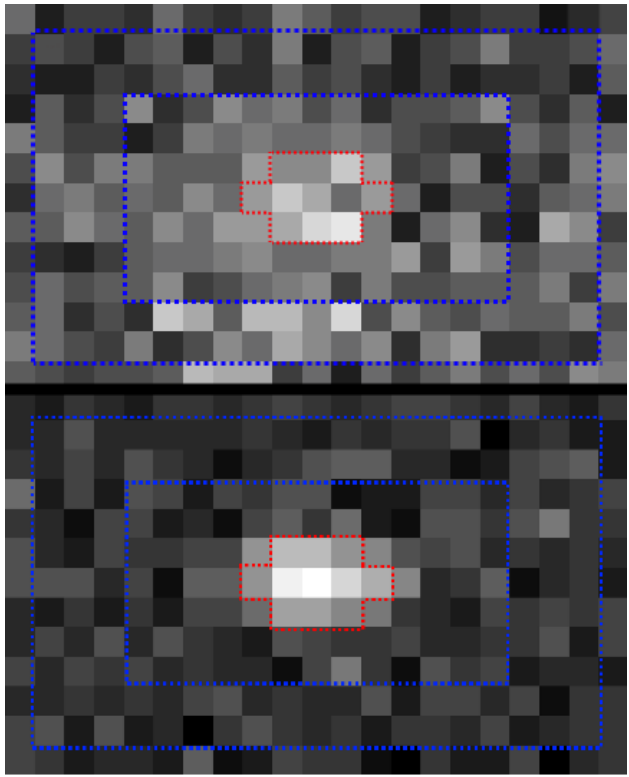}
       \caption[interior]{Representative postage stamps on the CCD
         detector with a signal found by the code (top) and a synthetic laser emission line inserted to test recall rate (bottom). The 2-D PSF of the synthetic line is
         measured from a representative spectrum in our survey, in
         this case having a spectral resolution FWHM of 4.15 pixels
         (left-right direction) and spatial profile (seeing) FWHM of
         2.18 pixels (0.83 arcsec, up-down direction), representative of average seeing at
         Keck 1 on Mauna Kea. The innermost outlined pixels represent the
         signal region, namely the pixels that contribute 95\% of the
         counts in the model PSF, spatially symmetrical about the
         central pixel. The outer, rectangular outlined area delineates the
         perimeter pixels used to calibrate the background noise RMS.}
     \label{fig:interior}
  \end{center}
\end{figure}

\begin{figure}
  \begin{center}
      \includegraphics[width=0.5\textwidth]{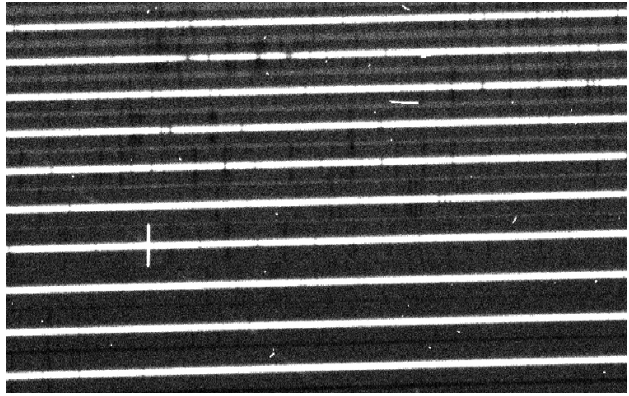}
       \caption[overlap]{Example of changing order separation on the
         middle CCD. Note that in the lower orders (higher wavelength)
         there is a faint gap of darkness, where neither neighboring
         order falls on this part of the CCD. As the orders get closer
         together, the sky light overlaps, creating a choppy
         texture. This is a frequent source of false positives if left
         unaddressed.}
     \label{fig:overlap}
  \end{center}
\end{figure}

To detect candidate laser emission lines in the ``sky'' region of the
CCD adjacent the stellar spectrum we constructed a routine that steps
pixel by pixel across all three CCDs searching for emission with the
requisite PSF properties described in Section~\ref{sec:method}.  We
construct ``postage stamps" with typical sizes of 11 x 19 pixels
(spatially and in the wavelength direction, respectively) centered on
each pixel, and we search for emission that meets specified criteria
within each stamp.  A representative postage stamp on the CCD is shown
in Figure \ref{fig:interior}.  We describe below the two null hypothesis ``gates''
by which we rule out prospective laser emission lines, involving a
requisite S/N ratio for the emission and goodness-of-fit criterion for
its PSF shape.

The first gate to be passed by the pixel and its postage-stamp
neighboring pixels tests the signal-to-noise ratio (S/N) of the
prospective laser line.  We establish a S/N threshold such that
fluctuations in the arrival of background photons are unlikely to rise
to a high S/N threshold.  The major sources of ``counts'' in a pixel are
bias, dark counts, readout noise, scattered light in the spectrometer, background sky
light, and laser emission, if any.

The bias, dark counts, and readout noise are easily determined, in
total, for each postage stamp by measuring the observed
fluctuations. This ``noise'' is measured by the RMS of the counts in the pixels around
the perimeter of each postage stamp, computed as follows:

\begin{equation}
RMS = \sqrt{\frac{\sum\limits_{i=1}^{N_{per}} (p_i-\mu)^2}{N_{per}}} ,
\label{eq:rms}
\end{equation}

 where the $p_i$ is the number of photons in the \it{i}\rm th pixel of the
 perimeter region, $N_{per}$ is the number of pixels in the perimeter
 region (typically 100-200 pixels, see Figure \ref{fig:interior}), and
 $\mu$ is the mean value of the number of photons in these
 pixels. We define $M$ to be the median value of the number of photons in the perimeter region.
We adopt the RMS in Eqn(1) as the noise in each pixel contributed by
the background fluctuations, bias, dark, readout noise, scattered light in the
spectrometer, and sky brightness.  The typical RMS is 2-3 photons per pixel.

This background noise is adopted for each pixel within the signal
region, the interior of the postage stamp, defined as the pixels in the model PSF containing 95\% of the
counts.  The number of pixels in the signal region ranges from 11 to
25 pixels depending on seeing. The total counts above the background
in the signal region are given by

\begin{equation}
I=\sum\limits_{i=1}^{N_{\rm pix}} (S_i-M) ,
\end{equation}

where the $S_i$ number of photons in the \it{i}\rm th pixel of the
signal region, $N_{\rm pix}$ is the number of pixels in the signal region
(typically 11-25 pixels).

We define a metric of the signal-to-noise ratio, ``S/N'', of the laser
emission signal as,

\begin{equation}
S/N = \frac{I}{\sqrt{N_{pix}\times RMS^2+I}}   , 
\label{eq:s/n}
\end{equation}

with the $RMS$ and $N_{\rm pix}$ defined as above. Hence our S/N is
the ratio of the total number of photons above background within the
signal region to the quadrature-summed noise expected in the signal
region, which has contributions from background fluctuations in each
pixel and from Poisson fluctuations in the signal. This is not a
precise signal-to-noise ratio, as we assume Gaussian distributed
background fluctuations.  As the total number of counts of the laser 
grows, loss of precision from the Gaussian background assumption 
is quickly dwarfed by poisson fluctuations in the signal. The goal is not to compute an
accurate signal-to-noise ratio of the prospective laser emission
signal. Rather it is to compute a quantity that is sufficiently close
to the signal-to-noise to identify and rank-order prospective laser
emission signals based on their standing above the noise, motivating
follow-up analysis and observations to assess reality of the signal.

We establish a second metric for the laser emission based on a
goodness-of-fit to the known 2-D PSF. We compute a reduced $\chi^2_{r}$
by fitting the photon counts within the postage stamp with the
previously determined 2-D PSF, serving as the model, with the only
free parameters being the position of the PSF.
We use the same pixels in the postage stamp that were used to compute
the S/N.  We scale our model PSF to have the same total photon counts as observed
in the postage stamp image. When computing reduced $\chi^{2}_{r}$, we
allow only the position of the PSF model to adjust to the observed
photon counts in their respective pixels.  We compute the reduced $\chi^{2}_{r}$ as

\begin{equation}
\chi^{2}_{r}=\frac{1}{\nu}\sum\limits_{i=1}^{N_{pix}} \frac{(S_i - E_i)^2}{\sigma_{i}^{2}}
\label{eq:chisquare}
\end{equation}

where $S_i, E_i,$ and $\sigma_{i}$ are the photon counts in the
$i$th pixel of postage stamp region and of the PSF model,
respectively, and the associated uncertainty for each pixel. Here, $\nu$ denotes the number of
degrees of freedom, in this case the number of pixels used in the sum,
minus three fitted parameters: the x-y position of the 2-D PSF model
and the total number of photon counts. Subpixel sampling would have allowed 
for greater precision, but was not employed in this algorithm. The reduced $\chi^2_r$
statistic is a sufficient goodness-of-fit metric for this application, as
the PSF model is locally linear in the pixel location \citep{Andrae2010}. 

\subsection{Setting Thresholds in S/N and $\chi^{2}$}
\label{sec:testing}

We adopt the detection criteria for laser emission by demanding that
the S/N be greater than some threshold and that $\chi^{2}_{r}$ for the
fit to the 2-D PSF be near unity, below some threshold.  That is, we
test the null hypothesis that there is no convincing laser emission
within each postage stamp.  The null hypothesis can be rejected if the
S/N resides above some threshold and the $\chi^2_r$ statistic is less
than some threshold, to be determined as follows.

To establish thresholds for S/N and $\chi^{2}_{r}$, we tested the ability
of our code to pick out simulated laser lines over the entire format
under many conditions. We adopted a strategy of injection and
recovery of laser emission. We randomly selected actual
observations and inserted synthetic laser lines composed of the 2-D
PSF with superimposed Poisson
fluctuations.  We injected fake laser emission ``dots'' having varying
S/N, different locations within the echelle spectrum format on the
CCD, and modest (10\%) variations in the 2-D PSF to simulate the
actual variability of the PSF from that adopted.
These injected fake laser emission 2-D PSFs revealed the distribution
of values of S/N and $\chi^2_r$ that the code detected.  We also
learned the number of false positives per image that results from
Poisson fluctuations.  

We found that our algorithm incurred a steep rise in false
positives for S/N less than 7, which corresponds to $\sim$100 photons
in the laser signal, after including noise from the background. Nearly
one false positive per image occurred with S/N at 7 or greater,
clearly indicating that we could not drop the S/N threshold to lower
values without incurring a rapid increase in false positives.

Injecting fake laser emission right at this threshold of S/N=7, our
tests showed 99\% of simulated lasers had $\chi^{2}_{r} \in
[0.7,2.0]$, and we had a false positive rate of approximately 0.1 per
observation, with each observation containing $~3\times 10^6$ pixels
considered as the center of the laser line \footnote{To those familiar
  with signal-to-noise thresholding, a false positive rate of 0.1 per
  3 million observations seems very high for signal-to-noise of
  7. This is an effect of the actual noise variance of all sources of
  noise within a postage stamp region, captured by the noise measure
  around the perimeter given in Equation \ref{eq:rms}.}.

We tested the effect of the code miscalibrating the 2-D PSF, and found
only a small rise in $\chi^{2}_{r}$ (see Table \ref{chired}).
Changing the simulated 2-D laser profile by as much as 30\% in both
dimensions still maintained, on average, $\chi^{2}_{r}$ below 2, for
S/N below 10. Similar tolerance was seen for simulated lasers centered
with sub-pixel precision. The worst case was a simulated laser
situated at the intersection of four pixels, in which cases the mean
in $\chi^2_r$ rose no higher than 1.2 for S/N below 10.

The discrepancies in $\chi^2_r$ caused by a miscalibrated 2-D PSF and
from untreated sub-pixel displacements in the laser line grow with
increasing S/N.  A laser line containing hundreds or thousands of
photons will have small associated fractional Poisson errors compared
to weak laser lines.  Any mismatch between the actual 2-D PSF and the
adopted model 2-D PSF can cause the value of reduced $\chi^2_{r}$ to
be much greater than unity simply due to the adoption of a model PSF
different from the actual PSF. We accounted for such errors in the
model PSF by artificially increasing the adopted noise (above Poisson)
associated with the number of photons per pixel. We set a lower bound
on the uncertainty in the noise at 20\% of the number of detected
photons per pixel.

Thus, in equation \ref{eq:chisquare}, we have $\sigma_i$ which is the
greater of $\sqrt{(0.2 \cdot S_i)^2 +RMS^2}$ and $\sqrt{S_i +RMS^2})$. While approximate
this floor in the uncertainty ensured that even an observed laser line
having twice the FWHM in both the spatial and wavelength directions as
the adopted PSF model along with $\rm S/N = 100$ would be detected.
For laser signals exceeding $\rm S/N = 10$, we set a new
$\chi^{2}_{r}$ threshold of $\chi^{2}_{r} < 10$. While this threshold
is high in light of the lower bound on $\sigma_i$, the relative
paucity of postage stamps with $\rm S/N > 10$ resulted in a manageable
number of candidates.  The advantage is that prospective laser emission
lines containing hundreds or thousands of photons will be identified
by our search algorithm, even if the FWHM of the PSF model is wrong by
a factor of two. That is, strong laser lines will be detected
independent of the integrity of the 2-D PSF.

\section{Results}
\label{sec:results}

Using the algorithm described in Section~\ref{sec:algorithm}, we
searched 2796 target stars for laser emission in the wavelength range,
3640-7890 \AA \, located in the ``sky'' region between 2-7 arcsec from
the star image along the entrance slit. Two criteria defined a
detection of a candidate laser line as described in
Section~\ref{sec:algorithm}.  The signal-to-noise of the photons
(above background noise sources) had to meet the
threshold, S/N$>$7.0, corresponding to $\sim$100 photons collected
within the 2-D PSF of the Keck-HIRES spectrometer. Also, the
goodness-of-fit to the 2-D PSF in the wavelength and spatial direction
had to meet $\chi^{2}_{r} <$ 2.5.

Using these thresholds, we found 10,155 candidate laser emission lines
with S/N between 7 and 10, and another 5449 candidates with S/N above
10. Of those with S/N $<$ 10, 3570 candidates had $\chi^{2}_{r} < 2.0$
indicating an acceptable match with the 2-D PSF. The ensemble of 15,604
laser-line candidates were subsequently analyzed by eye to rule out
those clearly inconsistent with a 2-D PSF shape.  We rejected
candidate laser lines that were clearly caused by instrumental
artifacts such as from the location of the laser line at the edge of a spectral order or
the CCD detector, flaws in the optics of the spectrometer
(i.e. internal reflections, 2nd-order light from cross disperser), flaws in the
CCD detector (``hot'' pixels), or bleeding of charge on the CCD.
We also rejected candidate laser lines as false positives due
to atmospheric effects such as night sky emission or 
to astrophysical effects (spatially extended nebulae). This rejection process left
a mere eight candidate lasers that survived, all being
consistent with the 2-D PSF and thus could be true laser emission from
unresolved sources.

\begin{figure}
  \begin{center}
      \includegraphics[width=0.5\textwidth]{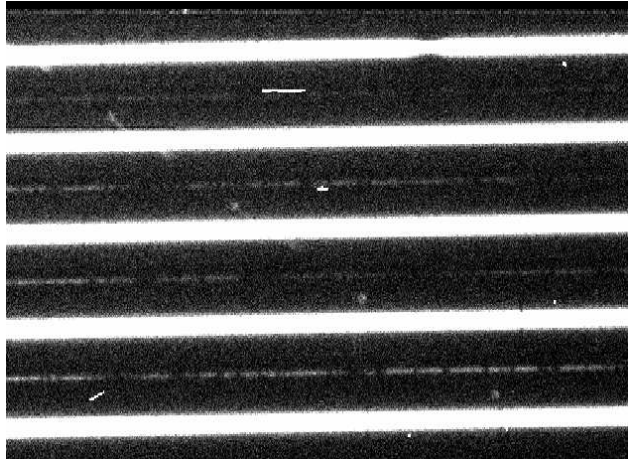}
       \caption[hip]{Diagonal pattern of dots on red CCD seen in
         HIP94931, which occurred in multiple exposures. The dashed
         stripe from top left to bottom right was displaced from night
         to night, and varies in intensity. This reflection pattern occasionally 
         occurred in other spectra. Note also the faint
         horizontal illumination between the spectral orders due to
         second-order diffraction from the cross disperser in the UV.}
     \label{fig:hip}
  \end{center}
\end{figure}

Each of these eight surviving laser candidates was analyzed again,
even more carefully by visually inspecting the raw CCD image.  We
looked for patterns of similar signals that were scattered spatially
in some linear or periodic fashion, either among neighboring spectral
orders or along the wavelength direction (within a few hundred
pixels). Such regular patterns are caused by internal reflections or
``ghosts'' within the spectrometer (see Figure 4).  These instrumental
effects can change from night to night due to slight repositioning of
the optics. Only by viewing
the candidate laser line with several levels of magnification (number
of pixels in field of view) was it possible to see the instrumental
pattern. In the case of HIP94931, of which we have more than 160
spectra, nearly PSF-shaped candidates appeared and disappeared from
observation to observation, and only by noting a set of collinear faint
dots from order to order was it possible to rule them out as candidate
lasers (see Figure \ref{fig:hip}). This is an example of one of the ``ghosts'' described above. In addition, though the area between
the orders was being ignored, in a few cases the second order UV light
from the cross disperser fell right between the orders in the red CCD
(at exactly 2x the wavelength). Our code was fooled by such 2nd-order
light in a few cases, as shown in Figure \ref{fig:hip}.

We ruled out seven of the eight laser candidates as artifacts, described above.
Also, for seven of the eight stars, we had obtained multiple spectra.  This allowed us to examine
the same pixel location (i.e. wavelength) to see if the laser emission
occurred in earlier or later exposures. Such examinations allowed us to identify
artifacts that were instrumental, even if recurring.
Finally, we also ruled out one laser candidate that was simply a spectrum of a
known planetary nebula angularly nearby the star being observed.

Thus, none of the eight candidates survived this more careful
examination. \it Therefore, our search of 14,380 spectra of 2796
different stars did not reveal any convincing evidence of laser
emission between 2 and 7 arcsec from the target star.\rm

 \section{Detection Thresholds of Laser Emission}
\label{sec:sens}

We now translate our detection thresholds of laser lines to the photon
fluxes at Earth, and to the luminosities of the lasers themselves,
that would have been detected.  Our imposed threshold of S/N$>$7 sets
the limit on the strengths of the signals we deem candidate laser
emission.  We use Equation \ref{eq:s/n} to compute the S/N ratio of the
prospective laser signal, accounting for both the Poisson fluctuations
in the candidate laser photons and the fluctuations in the background
noise.  This S/N threshold of 7 requires that $\sim$100 photons from
the laser ($I$ in Equation \ref{eq:s/n}) be acquired during the exposure to
be deemed a candidate laser signal.  The value of 100 photons varies
by $\approx10\%$ in different spectral regions and different
observations, depending on the specific background noise level from
scattered light and the sky.

We compute the corresponding laser flux at the Earth by considering
the effective collecting area of the Keck 1 telescope, 76 m$^2$,
and the efficiency of the telescope and HIRES spectrometer, 5\%,
which depends on the seeing and includes photon losses in the telescope optics, the
spectrometer, the entrance slit to the spectrometer, and the quantum
efficiency of the CCD detector \citep{Griest2010, Vogt1994}.  Thus, to detect
the threshold 100 photons at the CCD detector requires that 26.3 photons
per square meter fall on the primary mirror during the exposure.

We now consider exposure times that represent those used to obtain the
spectra in this study, ranging from 1 min (for Vmag = 7) to 45 min
(for Vmag = 13). (Note that the S/N per pixel in the stellar spectrum
achieved at Vmag = 13 is less than that achieved at Vmag=7.)  For
exposures of 1 min, the threshold photon flux from the laser required
to detect 100 photons is 0.44 photons m$^{-2}$ s$^{-1}$.  For
exposures of 45 min, the threshold photon flux to achieve 100
detected photons is 0.010 m$^{-2}$ s$^{-1}$, a remarkably low flux.

One may easily write these photon flux thresholds, $F$, in terms of
the power of the laser, $P$, the laser light frequency, $\nu$, and the
distance, $d$, from the laser to Earth:

\begin{equation}
F = \frac{P}{\pi h \nu \big[\theta d/2 \big]^2}
\end{equation}

Here, $\theta$ is the full opening angle of the diverging laser beam
(in radians).  As a benchmark example, we consider
``Keck-to-Keck'' laser transmission and reception.  We consider a 10m
diameter, diffraction-limited laser transmitter (in vacuum, outside
any atmosphere) detected by the Keck 10m telescope and HIRES
spectrometer. We adopt the normal Raleigh criterion for the beam divergence angle, $\theta = 1.22
\lambda/D$. Here, $\lambda$ is the laser wavelength and $D$ is the
diameter of the diffraction-limited laser emitter. A 
diffraction-limited laser itself has a narrower beam ``waist'' leading
to an intensity pattern having a characteristic angular beam size given by $\theta =
(2/\pi) \lambda/D$.  For $\lambda$ = 550 nm near the middle of
our wavelength sensitivity, and a 10m emitting aperture, the beam
divergence angle is $ 0\farcs014$. 

It is worth noting that such a laser concentrates its power into a
beam so narrow that its flux at Earth is 10$^{15}\times$ brighter than
an isotropic source of the same power and distance.  Moreover, its
monochromaticity delivers that energy within a narrow wavelength
range, $\sim$10$^{-4}$ of the wavelength width of the typical stellar
spectral energy distribution.  Thus, laser transmission from a
10-meter aperture achieves a detectability boost over a stellar
spectrum of a factor of 10$^{19}$.  For example, as Sun-like stars
have luminosities of 3.8 $\times 10^{26}$W, a diffraction-limited
laser with a power of only 4$\times$10$^7$W will deliver a light flux (at its
wavelength) outshining the host star. But, the laser must be aimed at
the Earth.

To continue the benchmark example, we adopt laser power similar to
that of existing laser guide star systems, which is 7 W at the Keck
Observatory (albeit at a wavelength of 588.995) nm.  We consider a
benchmark distance of 10 ly, representative of the nearest dozen
stars.  A 60s integration time (typical of our exposures of nearby
stars) yields $\sim$160 photons detected by our CCD, implying a S/N
ratio of 9.0-11.5 depending on seeing and local background noise
RMS. Thus, we would detect a 7 W laser located at a distance of 10
light years, as it would sit well above our detection threshold of S/N=7.

As mentioned in Section \ref{sec:HIRES_setup}, the exposure times of the two
major populations of target stars depended on visual magnitude. For
stars brighter than Vmag = 10, we stop exposures when the exposure
meter attains 250,000 ``counts'' (on an arbitrary scale),
representative of the nearby star target population. We stop exposures
for fainter stars, such as the {\it Kepler} stars, when the exposure
meter attains 60,000 counts.  As the majority of targets are main
sequence FGKM stars, the laser power from {\em Kepler} stars would need to
be about four times as high as from the nearby stars to be detected,
i.e., the greater (45 min) exposure times for the {\it Kepler} stars is
still $\sim$4x too small to make up for their greater distance.

We consider here the laser power required to reach a threshold S/N
ratio of 7, requiring 100 photons as previously discussed, for the two
representative populations of stars.  The average nearby star in our
survey resides at a distance of typically 30 pc = 100 ly and the
typical exposure time is 5 min.  For these nearby stars, the required
laser power for a threshold detection is $\sim$90 W.  The {\em Kepler}
target stars reside at a distance of $\sim$300 pc and the
typical exposure time is 45 min.  Thus for the {\it Kepler} stars, the
required laser power for a threshold detection is $\sim$1 kW.

Our initial pass at detecting laser lines suffered from an upper bound
in detectability at a S/N ratio of 400, due to saturation of the CCD
in some pixels at this exposure level.  We overcame this by carrying
out a second pass that permitting any high photon level, including
saturation.  No such saturated laser line candidates were found.

\section{Discussion}
\label{sec:Disc}

The power, beam dispersion, and distance to a continuous-wave laser
determine its detectability with the HIRES spectrometer on the Keck 1
telescope for the 2796 stars observed here.  A transmitting
extraterrestrial civilization could produce a detectable laser signal
by any combination of the size of their diffraction-limited
optics and the power of their laser.  Narrower laser beams concentrate
their intensity but obviously require more precise pointing for our
telescopes to intercept the beam, and laser tracking precision becomes
more important. The Keck-to-Keck case considered above demands laser
pointing accuracy of $\sim$10 milliarcseconds, set by the
diffraction-limited beam size, a pointing accuracy achieved with
current spaceborne telescopes and avionics.  Advanced civilizations
could presumably have similar, if not better, pointing ability and laser technology.

Our Keck-HIRES experiment achieves sensitivity to remarkably low
photon fluxes of 0.4 m$^{-2}$ s$^{-1}$ in a 1 minute exposure and
proportionally lower thresholds for longer exposures.  There are
many configurations of the laser source parameters that would allow
for detectability at Keck with required laser power at kilowatt
levels from distances over 1000 light years. At such distances,
extinction of optical light due to interstellar dust is only a few
tenths of a magnitude in V band, decreasing the laser fluxes at
Earth by no more than 50\%.  Thus extraterrestrial lasers of
kilowatt power easily permit detection from distances of 1000 light
years.

Such kW laser power is routinely achieved with present technology.
The Keck 1 Laser Guide Star AO system has a power of 7 W, only an
order of magnitude too weak to be detectable at the typical distances
of the nearest 1000 stars, even if the beam width were diffraction
limited. But many current lasers are far more powerful, easily
detectable if emitted from the distances of our 2796 target stars.
Commercially available YAG (yttrium aluminium garnet) lasers have a
power of 125 W operating at a wavelength of 532 nm, easily detected by
Keck-to-Keck transmission and reception \citep{Laserfabrik}. Similarly,
the US Navy has deployed 100 kW solid-state chemical iodine lasers
working in the near-IR for use on combat vessels, which, while not in
the wavelength band explored here, would be detectable at Earth with
even modest beam width and pointing accuracy\citep{lockheed2014}.  Considering another
civilization's technical advancement relative to our own during just
100-1000 years, the required laser power of kW levels seems within
expectation.

We may compare our current technique for laser detection with the past
searches for optical pulsed-lasers (OSETI).  Such searches are
sensitive to very short pulses consisting of several photons during a
nanosecond pulse time scale.  Their domain of strength over our method
is in the detection of isolated optical laser pulses that are briefly
(during a few ns or $\mu$s) brighter than the host star.  As a modern
example, the proposed NIROSETI experiment \citep{Maire2014,
  Wright2014_S} at the Lick Observatory 1-meter Nickel telescope will
be sensitive to 40 photons in the near-IR per square meter arriving in
a pulse of width 0.5 ns.  At Keck, a single such pulse falling on the
Keck telescope during an exposure yields $\sim$150 photons incident on
the HIRES CCD detector, a greater number of photons than with NIROSETI
due to the large aperture of the Keck telescope. The $\sim$150 photons
from just one ns pulse is easily detectable with our method, provided
the signal does not have to compete with the star's flux. If the
telescope picks up many such pulses, contained perhaps
in a train, the laser signal would be even more detectable
with our method.  In a five minute exposure, pulse cadences of Hz,
kHz, and MHz would strengthen the signal as detected by our method,
due to the multitude of photons.  

One caveat in our work bears emphasis.  Our examination of over 2796
stars for unresolved laser emission was done by avoiding examination
of the sky region within 2 arcsec of the star itself.  We explicitly
searched for laser signals that were resolved spatially from the
target star, given the typical seeing profile having FWHM of $\sim$1
arcsec.  The actual separation necessary to resolve the laser line
ranged from approximately 5 pixels (2 arcsec) from the center of the
stellar spectrum, to as many as 10 or 12 pixels (3-4 arcsec) on nights
of exceptionally poor seeing. For each spectrum of each star, the
spatial width of the spectrum determines the closest angular
separation from the star, typically 2-3 arcsec. Clearly this ``inner
working angle'' of 2-3 arcsec constrains the physical distance of any
detectable lasers from the target star, typically 10-100 AU for the
nearest stars.  Similarly we can detect lasers located 2000-7000 AU from the
{\it Kepler} stars.

This restriction on the distance of the laser from the star bears                                        
directly on the 10 light-year Keck-to-Keck thought experiment                                            
described above. An Earth-analog planet hosting a powerful laser                                         
located 1 AU from any target star would reside angularly within the                                      
inner working angle of all 2796 of our target stars, leaving such                                        
lasers undetected in this current search.  In effect, this current search is sensitive to                
lasers that are associated with technological constructs located many                                    
AUs from a star.   

In a following paper, we will report on a search for laser emission
coming from {\em within} the current 2 arcsec inner working angle, i.e.
spatially coincident with the stars themselves.  The corresponding flux
thresholds for detection of lasers will be higher (worse) by roughly
two orders of magnitude due to the Poisson fluctuations of the
$\sim$10$^4$ photons per pixel.  Similar searches of stellar spectra
for laser emission can also be carried out on other existing
spectroscopic surveys, such as the Sloan Digital Sky Survey. The
technique will work equally well in the infrared, with the added
bonus of lessened extinction from dust in the Galaxy.  Such extant
spectroscopy surveys may lack the resolution 
to demonstrate the monochromatic nature of the laser emission. However, 
follow-up observations of compelling candidates could be 
performed with high resolution spectroscopy to support or reject the
interpretation of laser emission.  A robust toolset able to
search for laser lines in a variety of different telescope spectra
would be a useful addition to the SETI community at large.

\section{Acknowledgments }

\indent 
We thank the John Templeton Foundation for funding this research. 
We thank John Johnson and Andrew Howard for the Keck 
Telescope time awarded to CalTech and to the University 
of Hawaii and the Institute for Astronomy (IfA) allowing many of 
these spectra to be obtained.  We thank Howard Isaacson for
enormous help with the observations, reduction, and organizing target
lists.  G. Marcy, the Alberts Chair at UC Berkeley, would like to
thank Marilyn and Watson Alberts for funding and support that made this
research possible. The spectra presented herein were obtained 
at the W. M. Keck Observatory, which is operated as a
scientific partnership among the California Institute of Technology,
the University of California, and the National Aeronautics and Space
Administration. The University of California Observatories (UCO)
provided key support of HIRES and other instrumentation at the Keck
Observatory.  We thank the state of California and NASA for funding
much of the operating costs of the Keck Observatory.  The construction
of the Keck Observatory was made possible by the generous financial
support of the W. M. Keck Foundation. We thank the many observers who contributed to the
measurements reported here. We gratefully acknowledge the efforts and
dedication of the Keck Observatory staff, especially Scott Dahm, Hien
Tran, Grant Hill and Greg Doppmann for support of HIRES, and Greg
Wirth, Bob Goodrich, and Bob Kibrick for support of remote
observing. This work made use of the SIMBAD database (operated at CDS,
Strasbourg, France) and NASA's Astrophysics Data System Bibliographic
Services. This research has made use of the Kepler Community Follow-up
Observing Program (CFOP) database and the NASA Exoplanet Archive,
which is operated by the California Institute of Technology, under
contract with the National Aeronautics and Space Administration under
the Exoplanet Exploration Program. The authors wish to extend special
thanks to those of Hawai`ian ancestry on whose sacred mountain of
Mauna Kea we are privileged to be guests. Without their generous
hospitality, the Keck observations presented herein would not have
been possible. We thank Dan Werthimer, Andrew Siemion, Jill Tarter, 
Frank Drake, Jason Wright, Lucianne Walkowicz, Shelley Wright, 
John Gertz, Andy Fraknoi, David Brin, Charlie Townes, Mike Garrett, 
Amy Reines, and Phil Lubin for valuable conversations about optical 
SETI. We also thank the ``Berkeley SETI Research Center" (BSRC) and 
the ``Foundation for Investing in Research on SETI Science and 
Technology'' (FIRSST) for ideas and support toward the future of 
SETI research.

\bibliographystyle{apj}

\begin{deluxetable}{ccc}
\tabletypesize{\small}
\tablenum{1}
\tablecaption{Mean $\chi^{2}_{r}$ Values for Simulated Lasers, S/N of 10}
\label{chired}
\tablewidth{0pt}
\tablehead{
\colhead{Error in FWHM$_{spat}$ }    & \colhead{Error in FWHM$_\lambda$}      & \colhead{Mean $\chi^{2}_{r}$}}
\startdata
0.0	&0.0 & 1.1(2)\\
$\pm$20\%	&0.0 & 1.3(3)\\
0.0	&$\pm$20\% & 1.2(3)\\
$\pm$20\%&$\pm$20\%&1.5(3)\\
$\pm$30\%&$\pm$30\%&1.7(4)
\enddata
\tablecomments{Mean values of $\chi^{2}_{r}$ from lasers with miscalibrated PSF. The expected error in the PSF is an order of magnitude lower than the limits considered here. }
\end{deluxetable}

\enddocument